# Two-dimensional *spinodal* interface in one-step grown graphene-molybdenum carbide heterostructures

Jia-Bin Qiao[1], Yue Gong[2,4], Haiwen Liu[1], Jin-An Shi[2,4], Lin Gu[2,3,4] and Lin He[1,*]

[1]Center for Advanced Quantum Studies, Department of Physics, Beijing Normal University, Beijing, 100875, People's Republic of China
[2]Beijing National Laboratory for Condensed Matter Physics, Institute of Physics, Chinese Academy of Sciences, Beijing 100190, China
[3]Collaborative Innovation Center of Quantum Matter, Beijing 100190, China
[4]School of Physical Sciences, University of Chinese Academy of Sciences, Beijing 100190, China
[*]Correspondence and requests for materials should be addressed to L.H. (E-mail: helin@bnu.edu.cn).

**Heterostructures made by stacking different materials on top of each other are expected to exhibit unusual properties and new phenomena. Interface of the heterostructures plays a vital role in determining their properties. Here, we report the observation of a two-dimensional (2D) *spinodal* interface in graphene-molybdenum carbide ($\alpha$-Mo$_2$C) heterostructures, which arises from *spinodal decomposition* occurring at the heterointerface, by using scanning tunneling microscopy. Our experiment demonstrates that the 2D *spinodal* interface modulates graphene into whispering gallery resonant networks filled with quasi-bound states of massless Dirac fermions. Moreover, below the superconducting transition temperature of the underlying $\alpha$-Mo$_2$C, the 2D *spinodal* interface behaves as disorders, resulting in the breakdown of the proximity-induced superconductivity in graphene. Our result sheds new light on tuning properties of heterostructures based on interface engineering.**

The interface, simply produced by assembling two materials together, creates complicated circumstances differing drastically from the bulk [1,2]. Strain relaxation [3], charge distribution [4] and interactions [5-7] in interface of heterostructures can result in novel quantum phenomena and properties not found in their constituents [5-8]. For example, the interface in $LaAlO_3/SrTiO_3$ heterostructures behaves as 2D electron gas [9], possessing both superconductivity and ferromagnetism [10,11]; monolayer FeSe on $SrTiO_3$ exhibits superconductivity at much higher temperature than its bulk [8]; the interfacial moiré potential in graphene/h-BN heterostructures results in the formation of secondary Dirac points and the Hofstadter's butterfly pattern [5-7]. Obviously, the realization of the novel properties is intimately linked to the fabrication of well-defined heterostructures with high-quality interfaces.

In this work, we systematically studied one-step grown graphene/α-$Mo_2C$ heterostructures and unexpectedly observed novel 2D *spinodal* interfaces, originating from dispersive Mo-C atomic gas *via spinodal decomposition* [12] occurring at the interface. *Via* scanning tunneling microscopy (STM) measurement, we demonstrated that the 2D *spinodal* interface strongly affects electronic properties of the heterostructures. It not only introduces confining potentials into graphene and modulates it into 2D whispering gallery resonant networks occupied by quasi-bound states of massless Dirac fermions [13], but also behaves as disorders, which result in the breakdown of proximity-induced superconductivity in graphene.

**Results**

**Synthesis and characterization of graphene/α-$Mo_2C$ heterostructures**

In our experiment, we synthesized large-scale high-quality graphene/α-Mo$_2$C heterostructures *via* one-step chemical vapor deposition method [14]. In short, diffused Mo atoms across molten thin Cu foil firstly react with C atoms dissociated from methane (CH$_4$) to form well-defined 2D layered α-Mo$_2$C crystals [15], as shown in Fig. 1a and 1b, and then the preformed carbides proceed to catalytic synthesis of graphene on their surface to generate the heterostructure [14], as shown in Fig. 1c. The formation of well-defined graphene/α-Mo$_2$C heterostructures have been confirmed by various characterizations, such as high-resolution scanning transmission electron microscopy (HR-STEM) and STM (see more characterizations in Supplementary Fig. S2-S4 and Note 2 [16]). Fig. 1d exhibits a representative selective area electron diffraction (SAED) pattern along the [100] zone axis of the heterostructures. The coexistence of two types of ED patterns from monolayer graphene and α-Mo$_2$C crystal supports the formation of the heterostructures. Here we notice that there is a rotated angle (~ 11°) between graphene and α-Mo$_2$C lattices, which is consistent with our previous experimental result [14]. The displacement of the heterostructures may arise from the unique one-step-growth mechanism. In the synthesis process, the graphene grows continuously with a plethora of C source relative to Mo atoms following the carbide growth. Due to lattice mismatch between the two crystals, accompanied with the 2D *spinodal decomposition* emerging at the interface, epitaxial graphene sheets may rotate relative to underlying α-Mo$_2$C crystals for a thermally stable stacking mode in the one-step-grown heterostructures. Fig. 1e shows a typical atomic-scale STM image of the heterostructure surface, where a hexagonal honeycomb lattice of monolayer graphene is clearly observed. In Fig. 1f and 1g, we show the atomic-resolution high-angle annular dark-field (HAADF)-STEM images of the plane and section views of the heterostructures, where the two sets of lattices match well with α-Mo$_2$C (100) and (010) surfaces, respectively. All the characterizations validate the realization of well-defined graphene/α-Mo$_2$C heterostructures. Direct synthesis of heterostructure largely prevents interfaces from exotic contamination and is believed to be an effective method to create high-quality interface. However, a typical HAADF-STEM image taken around the interface, as shown in Fig. 1h, shows a nanoscale blurred interfacial region between top graphene (not obviously distinguished mostly due to the low scattering cross section of C comparing with Mo) and underlying α-Mo$_2$C crystals, unveiling the interface in the one-step grown heterostructures is not as perfect as expected.

**Atomic topography of the 2D *spinodal* interface**

To further characterize the interface, we performed STM measurements on the heterostructures. Although the interface is coated by graphene, the single-atom layer provides an excellent looking window rather than an obstacle for the visualization of the underlying interface in STM studies [17]. The microscopic investigation of the interface reveals the presence of four types of topographic morphologies. The first morphology shows dispersive island-like patterns with a size ranging from several to tens of nanometers and a height of ~ 0.5 nm, as shown in Fig. 2a. The second one exhibits isolated vacancy-island patterns and the vacancy islands show similar size and height as the islands of the first type, as shown in Fig. 2b. In contrast to the said two morphologies, the third one displays weird labyrinth-like interconnected patterns with width around several nanometers and a lower height (~ 0.2 nm), as shown in Fig. 2c and 2g. Fig. 2d and 2h show the last morphology featured by a plain-like morphology in a large range. Intriguingly, however, a close-up of the "flat" structure unveils that it is not flat but a fractal pattern of the third one, i.e., self-similar pattern of labyrinthine interconnection in sub-nano scale (less than 1 nm in width and tens of picometer in height). Figures 2e-2h show typical profile lines of the four types of topographic morphologies. In our experiment, we found that the said four types of topographic morphologies can be observed on all the studied heterostructures, and the top graphene sheet is perfectly continuous and almost free of defects across the whole area for all the studied topographic morphologies.

**Theoretical simulation of the 2D *spinodal* interface**

Our analysis indicates that the distinctive topographic patterns observed in our experiment arise from the formation of 2D interfacial dispersive Mo-C atomic gas by *spinodal decomposition* [12] occurring at the heterointerface and the volume fraction of the Mo-C atomic gas plays a vital role in determining the final topographic patterns. In the post-carbide-growth stage, diffused Mo atoms are close to saturate while supply of C atom is still increasing. Sufficient C atoms proceed to self-assembly for graphene growth assisted with graphitic catalysis of the preformed carbides, leaving a gas of randomly dispersive Mo atoms embedded with isolated C atoms in the heterointerface [18]. The interfacial Mo-C atomic gas, as a new structural phase, separated from the bulk in a *spinodal* fashion, termed *spinodal decomposition*, which is analogous to a gas of Au adatoms on Au (111) surface [19].The heterogeneous system (i.e., the interfacial

Mo-C atomic gas on α-Mo$_2$C crystals) commonly undergoes phase separation and hence exhibits remarkable structural evolutions, which could be described by a conventional binary phase diagram. In the binary phase diagram, the unstable region is defined by the *spinodal* curve [20,21]. When a system crosses the curve into the unstable region, phase separation occurs spontaneously without passing through a nucleation and growth stage, corresponding to the *nucleation and growth* mechanism. As pointed out by Gibbs, upon penetrating deeply enough into this region, the potential barrier against the formation of stable nuclei disappears and the system turns into an unstable stage. Due to the lack of any activation barrier, the spontaneous and instantaneous, i.e., *spinodal*, decomposition of the system initiates with considerable density fluctuations of a certain wavelength growing exponentially with time. The procedure is known as *spinodal decomposition* (see Supplementary Fig. S1a [16]). The initial volume fraction of the component dramatically determines the ultimate density distribution, i.e., morphology, of the system undergoing *spinodal decomposition*[18,21] (see more details in Supplementary Note 1 [16]). To further understand the effect, we carried out a theoretical simulation based on Cahn-Hilliard (CH) model [23]. *Via* an unconditionally stable Fourier-spectral method, the CH equation in 2D space could be transformed into discrete grid points by using the linearly stabilized splitting scheme,

$$\frac{\phi_{ij}^{k+1} - \phi_{ij}^{k}}{\Delta t} = \Delta \left( 2\phi_{ij}^{k+1} - \varepsilon^2 \Delta \phi_{ij}^{k+1} + f\left(\phi_{ij}^{k}\right) \right). \qquad (1)$$

Here $f(\phi) = \phi - \phi^3$, $\phi = c_B - c_A$ is a function as time $t$, where $c_A$ and $c_B$ denote the compositions (or volume fractions) of components $A$ and $B$, respectively. $\varepsilon^2$ is a constant called the gradient energy coefficient, which is related to the interfacial energy.

Then we could numerically solve the equation (1) to perform the simulation of morphological patterns during *spinodal decomposition*. In the simulation, the initial condition is taken to be $\phi(x,y,0) = \alpha\left(\mathrm{rand}(x,y) + \beta\right)$, where rand$(x, y)$ is a random number between -1 and 1, providing a random density perturbation (background). One can adjust parameters $\alpha$ and $\beta$ to change the initial volume fraction (0 to 100%) of $A$ (or $B$). When the volume fraction of the Mo-C atomic gas is less than and more than 50% in the heterointerfacial system, the morphologies are characterized by the island and vacancy-island patterns respectively, as shown in Fig. 2i and 2j (see also Supplementary Fig. S1b-d and h-j [16]). For the case that the volume fraction is close to 50%, the nanoscale labyrinthine interconnected patterns evolve, as shown in Fig. 2k (see also Supplementary Fig. S1f and S1g [16]). Moreover, in early stage of the *spinodal decomposition* when the gas with ~ 50% volume fraction, we obtained sub-nanoscale labyrinth-like fractal patterns in our simulation, as shown in Fig. 2l (see also Supplementary Fig. S1e [16]). Note that the fundamental features of the distinctive morphologies observed in experiments are reproduced well by the simulated topography (see Supplementary Note 1 [16]). All the results confirm the formation of 2D *spinodal* interface between graphene and α-Mo$_2$C in the one-step grown heterostructures. Here we should point out that *spinodal decomposition* occurs before the graphene growth (see Supplementary Fig. S12 [16]) and graphene plays a negligible role in the formation of the 2D *spinodal* interface.

**Spinodal-modulated whispering gallery resonant networks occupied by quasi-bound states of massless Dirac fermions**

Now we turn our attention to the effect of the 2D *spinodal* interface on the electronic properties of the heterostructures. The 2D *spinodal* interface can locally affect doping of graphene and generate complex network of *n-p-n* and *p-n-p* junctions in it [13,24-27]. Because of the Klein tunneling in graphene [28], i.e., the unusual anisotropic transmission of the massless Dirac fermions at the *p-n* junction boundaries, these complex network of the *p-n* junctions are expected to confine charge carries in graphene to form quasi-bound states [13,24,25]. Figure 3 shows a typical experimental result, which demonstrates the formation of quasi-bound states in both the *n-p-n* and *p-n-p* junctions generated by the 2D *spinodal* interface. Fig. 3a displays a typical *spinodal* topography decorated with the island-like patterns. Our high-resolution STM measurements, as shown in Fig. 3b, demonstrate that the graphene sheet is perfectly continuous and free of any distortion or defect in the studied region. The quasi-ellipse-shaped island in Fig. 3b generates closed *p-n* junctions in graphene. Our spatial-resolved spectroscopic (STS) measurements at 4.2 K around this region, as shown in Fig. 3c, show a series of discrete resonances appearing in the tunneling conductance both inside and outside the quasi-ellipse-shaped island. These resonances are clearly evidence of the existence of quasi-bound states confined in the junctions. For a *n-p-n* junction, charge carriers below the Dirac point of the *p*-doped region are temporarily trapped to form the quasi-bound states. Whereas, massless Dirac fermions above the Dirac point of the *n*-doped region in a *p-n-p* junction are expected to be temporarily confined. Such results are demonstrated explicitly in our experiment (Fig. 3c). The existence of confined quasi-bound states was further confirmed by carrying out STS mapping measurements, as shown in Fig. 3d-3f. The STS maps at different sample biases reflect spatial distribution of the electronic modes, i.e., local density-of-state (LDOS), at different energies of the sample. At high energy ($\sim 0.4$ eV), LDOS is evenly distributed in all the area because of the absence of any confined modes, as shown in Fig. 3d. However, at low positive energy ($\sim 0.1$ eV) and negative energy ($\sim$ -0.1 eV), the STS maps display strong intensity both inside and outside the island region, as shown in Fig. 3e and 3f respectively. Similar phenomena of the quasi-bound states have been widely observed in various *spinodal* patterns (see Supplementary Fig. S7 and S8 [16]). All the results reveal that the 2D *spinodal* interface gives rise to the doping effect, and strikingly provokes the transformation from pristine graphene sheets into whispering gallery resonant networks filled with quasi-bound states. Of cause, the strong interaction between Mo and graphene [29,30] also can lead to the electronic hybridization and band energy shifts [30-34], which significantly affect electron scattering, contributing to the realization and modulation of quasi-bound states as well.

**Spinodal-driven breakdown of proximity-induced superconductivity in graphene**

The 2D *spinodal* interface not only generates complex network of *p-n* junctions in graphene, but also strongly affects proximity-induced superconductivity in graphene [14]. We carried out STM/STS measurements on the heterostructures at around 400 mK (below the superconducting transition temperature $T_c \sim 4$ K, see Supplementary Fig. S11a [16]), and discovered two distinct tunneling spectra acquired on different *spinodal* morphologies, as shown in Fig. 4. For the structure shown in Fig. 4a (type I), the local interface exhibits weak *spinodal* phase separation and we observe an unusual deformed hexagonal lattice, as evidenced by the fast Fourier transform (FFT) of the topographic image. The deformed hexagonal lattice is attributed to the defective (2×2) reconstruction of the Mo-C atomic gas on α-Mo$_2$C (100) surface (lattice distance $a_{2\times2}$ ~ $2a_0$ ~ 0.6 nm, see Supplementary Fig. S2b [16]). Our STS measurements observe a typical superconducting gap, as the hallmark of the proximity-induced superconductivity, in the type I structure (Fig. 4c). However, for the type II structure (Fig. 4b), which shows a typical topographic image of nanoscale labyrinthine interconnected patterns, there is no superconducting gap in the tunneling spectrum, as shown in Fig. 4c, indicating that the proximity-induced superconductivity in graphene is completely destroyed. The non-superconducting feature is universal in this region and independent of the height fluctuation of the nanoscale labyrinth-like *spinodal* morphologies (see Supplementary Fig. S10 [16]). Such a phenomenon is attributed to the effects of the 2D *spinodal* interface, which behaves as disorder that remarkably enhances quantum fluctuations and suppresses quantum condensations [35,36], leading to the collapse and even absence of the proximity-induced superconductivity in graphene.

In conclusion, we directly observed the 2D *spinodal* interface, originating from *spinodal decomposition* occurring at the heterointerface, in the one-step grown graphene/α-Mo$_2$C heterostructures. The 2D *spinodal* interface not only creates the graphene-based whispering gallery resonant networks filled with quasi-bound states of massless Dirac fermions, but also behaves as disorder, which completely destroys the proximity-induced superconductivity in graphene. Our results shed new light on the interface of the atomic-scale synthesized heterostructures, as well as offer an appealing ground for further understanding of interface physics and design of functional heterostructure devices based on interface engineering.

**Acknowledgements**

This work was supported by the National Natural Science Foundation of China (Grant Nos. 11674029, 11422430, 11374035), the National Basic Research Program of China (Grants Nos. 2014CB920903, 2013CBA01603). L.H. also acknowledges support from the National Program for Support of Top-notch Young Professionals, support from "the Fundamental Research Funds for the Central Universities", and support from "Chang Jiang Scholars Program".

**Figures**

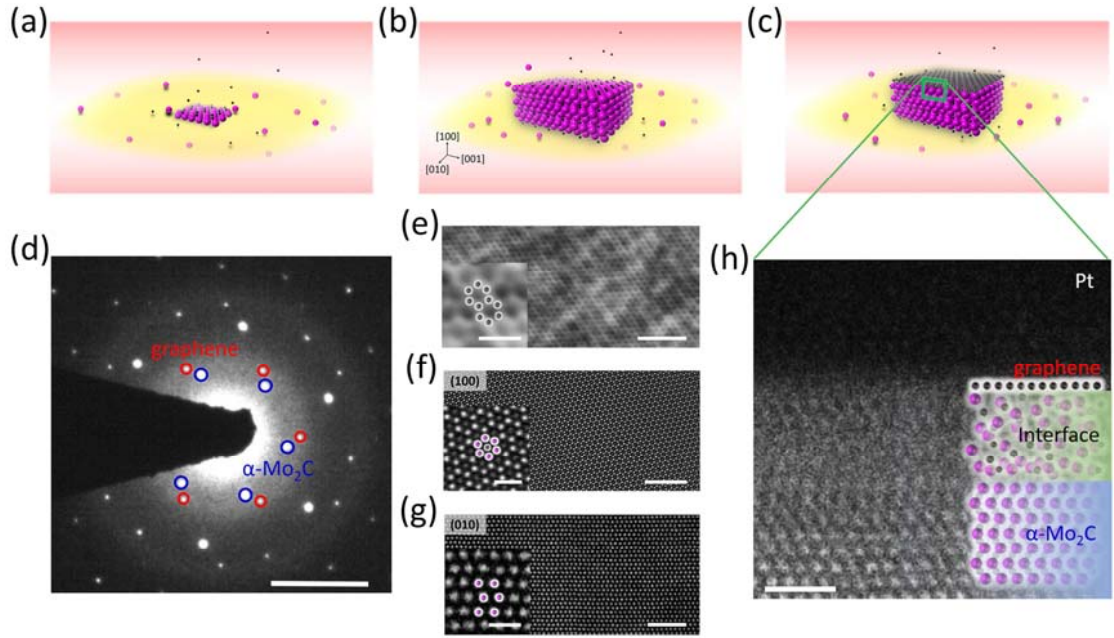

**Figure 1. Characterization of the one-step grown graphene/α-Mo₂C heterostructures. (a)** Mo (purple balls) and C (black balls) atoms initiate the preformation of α-Mo₂C. **(b)** α-Mo₂C crystal gradually grows up for synthesis of graphene. **(c)** The formation of the well-defined graphene/α-Mo₂C heterostructures. **(d)** A SAED pattern along the [100] zone axis, where some red (blue) spots belonging to graphene (α-Mo₂C) are circled. Scale bar, 5 nm⁻¹. **(e)** A typical STM image exhibiting hexagonal honeycomb lattice of monolayer graphene [marked by the black balls, inset in **(e)**], revealing the graphene on the heterostructure surface. **(f)** and **(g)**, Atomic-resolved HAADF-STEM images of the heterostructure surface parallel to the (100) facet and cross section parallel to the (010) facet, respectively, confirming the nature of underlying α-Mo₂C crytals where the center C atom (gray spot, marked by the black ball) is surrounded by six Mo atoms (white spots, marked by the purple balls) in a hcp arrangement in the (100) face and the Mo layers are arranged in AB-stacking form in the (010) face [inset in **(f)** and **(g)**]. Scale bars, **(e)-(g)**, 2 nm; insets, 0.5 nm. **(h)** A HAADF-STEM image taken around the interface of the heterostructures (with Pt coating as a protection layer). The blurred interfacial region corresponds to the 2D *spinodal* interface where Mo and C atoms are randomly distributed into the *spinodal* patterns. Scale bar, 1 nm.

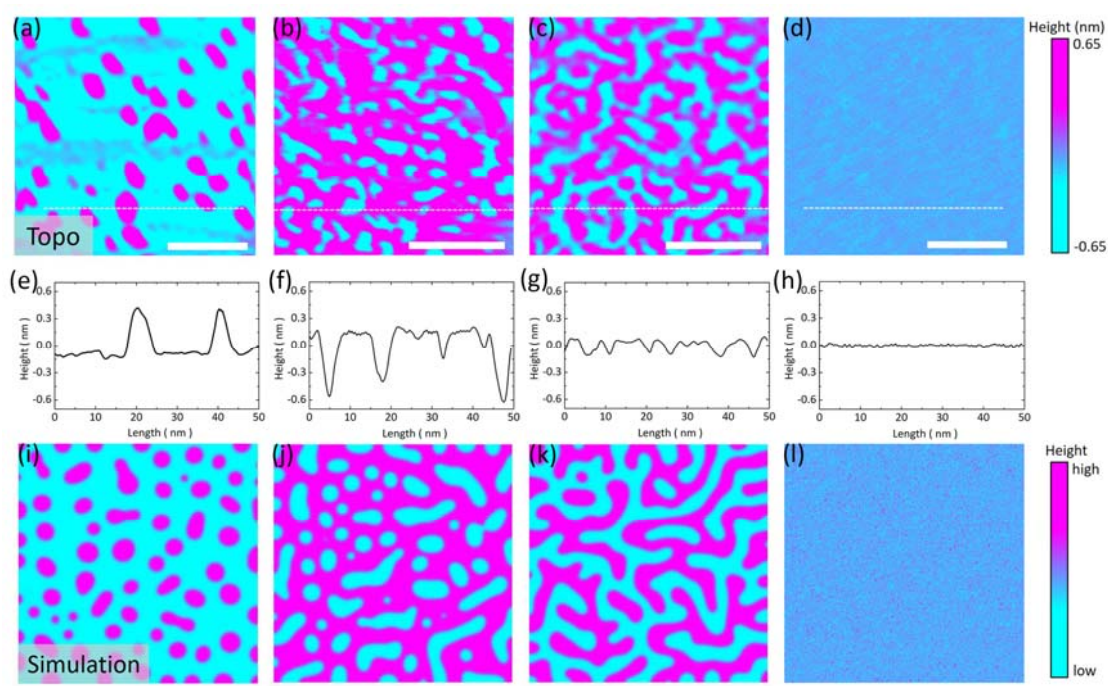

**Figure 2. Visualization and simulation of the 2D *spinodal* interface**, **(a)-(d)** Typical STM images exhibiting the island-shaped and vacancy-island-shaped patterns, nanoscale labyrinthine interconnected patterns and sub-nanoscale fractal patterns, respectively. The experimental parameters are: **(a)** $V_S$ = -100 mV, $I$ = 0.1 nA; **(b)** $V_S$ = -750 mV, $I$ = 0.2 nA; **(c)** $V_S$ = 600 mV, $I$ =0.1 nA; **(d)** $V_S$ = 400 mV, $I$ = 0.1 nA. Scale bars, **(a)-(d)**, 20 nm. **(e)-(h)** Height fluctuations along the white dashed lines in **(a)-(d)**, respectively. **(i)-(l)** Theoretical simulation on the *spinodal* topography with, **(i)** low initial volume fraction of Mo-C gaseous phase ($c$ < 50 %); **(j)** high volume fraction ($c$ > 50 %); **(k)** nearly half volume fraction ($c$ ~ 50 %); **(l)** $c$ ~ 50% in early stage of *spinodal*

*decomposition*.

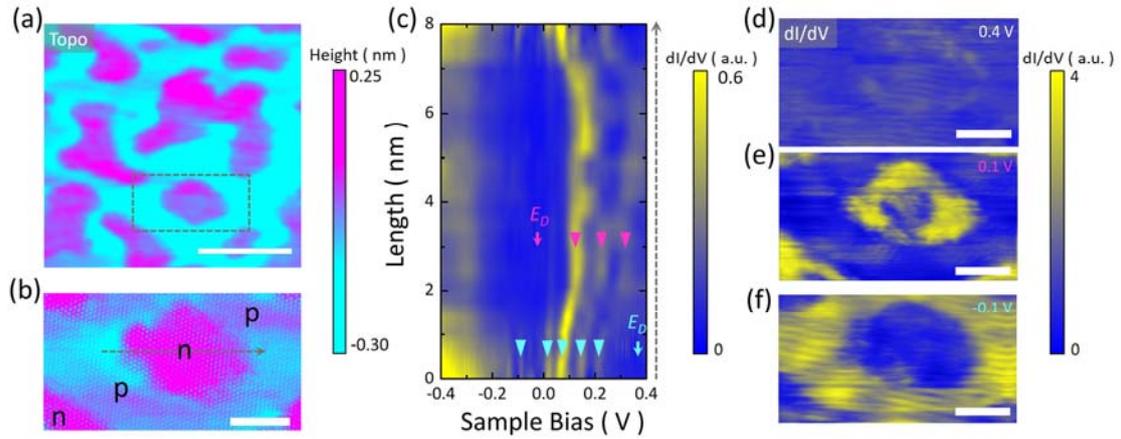

**Figure 3. Optoelectronic fiber network with whispering gallery modes. (a)** A large-scale STM image ($V_S$ = 800 mV, $I$ = 0.1 nA) of the island-shaped *spinodal* morphology. Scale bar, 10 nm. **(b)** A close-up ($V_S$ = -300 mV, $I$ = 0.1 nA) of an island pattern outlined by the black dashed rectangle in panel **(a)**. Scale bar, 3 nm. **(c)** Spatial resolved STS spectra measured along the black dashed arrow in panel **(b)**. The magenta (cyan) inverted triangles denote the modes belonging to region inside (outside) the island, respectively. The magenta (cyan) arrow labels the local Dirac point, indicating the *n* (*p*)-doped inside (outside) the island, respectively. **(d)-(f)**, STS maps measured around the same island at different sample biases, $V_S$ = 0.4 V **(d)**, 0.1 V **(e)** and -0.1V **(f)**.

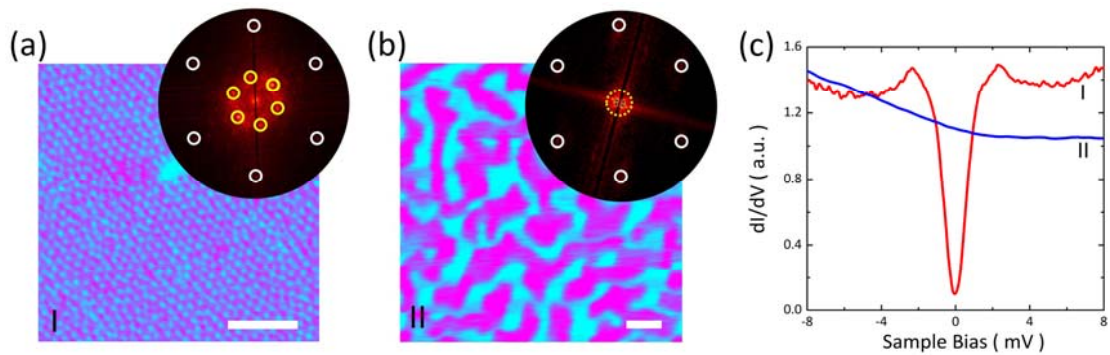

**Figure 4. Breakdown of proximity-induced superconductivity. (a)-(b)** Two typical STM images of the *spinodal* morphologies. The experimental parameters: $V_S$ = 1 V, $I$ = 0.1 nA (Topo I); $V_S$ = 800 mV, $I$ = 0.1 nA (II). Scale bar, 5 nm. The insets are the FFT images, where the outer six white circles denote the Bragg spots of the graphene reciprocal lattice, while the inner six yellow circles denote the Bragg spots of the reconstructed superlattice in Topo I and the inner yellow dashed circle outlines the disk-like area arising from the labyrinthine *spinodal* pattern in Topo II, respectively. **(c)** Tunneling spectra taken in Topo I and II, respectively.